\documentclass[aps,prd,eqsecnum]{revtex4}
\usepackage{graphicx}
\font\cst=cmr10 scaled \magstep4

\def\sqr#1#2{{\vcenter{\vbox{\hrule height.#2pt\hbox{\vrule
width.#2pt height#1pt \kern#1pt\vrule width.#2pt}\hrule height.#2pt}}}}
\def\square{\mathchoice\sqr54\sqr54\sqr{2.1}3\sqr{1.5}3} 

\begin{document}
\thispagestyle{empty}
%\vglue 1.5cm
\centerline{\cst  Newton's law}
\bigskip
\centerline{\cst on an Einstein ``Gauss-Bonnet" brane}
\vskip 1.5 true cm

\centerline{\large
Nathalie Deruelle$^{1,2}$ and Misao Sasaki$^{1}$}
\vskip 0.5 cm
\centerline{$^1$\it Yukawa Institute for Theoretical Physics}
\centerline{\it Kyoto University, Kyoto 606-8502, Japan}
\bigskip

\centerline{$^2$\it  Institut d'Astrophysique de Paris,}
\centerline{\it GReCO, FRE 2435 du CNRS,}
\centerline{\it 98 bis Boulevard  Arago, 75014 Paris, France}

\bigskip
\centerline{June 6th 2003}

\vskip1cm

\begin{center}
{\bf Abstract}
\end{center}
\bigskip

\noindent
It is known that Newton's law of gravity holds asymptotically on a flat ``brane" 
embedded in an anti-de Sitter ``bulk"~; this was shown not only when gravity in the
bulk is described by Einstein's theory but also in  Einstein  ``Lanczos Lovelock
 Gauss-Bonnet"'s theory.  We give here the expressions for the corrections to
 Newton's potential in
both theories, {\sl in analytic form and valid for all distances}. We find that in
 Einstein's theory the transition from the
$1/r$ behaviour at small $r$ to the $1/r^2$ one at large $r$ is quite slow. 
In the Einstein Gauss-Bonnet case on the other hand,  we find that the correction
 to Newton's potential
can be small for {\sl all} $r$. Hence, Einstein Gauss-Bonnet equations in the
 bulk (rather than simply Einstein's) induce on the brane a better approximation
 to Newton's law.

\section{Introduction}

There has been of late an increasing interest for gravity theories formulated
within spacetimes with one ``large" extra dimension, that is, the idea that
our universe may be a four dimensional singular hypersurface, or ``brane",
embedded in a five dimensional spacetime, or ``bulk". The Randall-Sundrum
scenario~\cite{RanSun99b}, where our universe is
a four dimensional quasi-Minkowskian edge of a double-sided perturbed anti-de
Sitter spacetime which satisfies the five dimensional Einstein equations, was
the first explicit model where the linearized {\sl four} dimensional Einstein
equations were found to hold on the brane, at least far from the  sources,
Newton's potential being corrected by
a small  $1/r^2$ term. This claim was thoroughly analyzed and the corrections
to Newton's law  exactly calculated, either in the small $r$ limit where it was
shown that they
diverged in $1/r$ or in the large distance limit~\cite{GarTan99}. The analysis was
then extended to the case when gravity in the bulk is described by the Einstein
Gauss-Bonnet equations, and it was found that, in that theory as well,
Newton's law  also held on the brane at large distance,\cite{KimLee00,Neu01}
with various results on the exact $1/r^2$ corrections, due to more or less
careful treatments of ``brane bending" and conflicting views on the boundary
conditions to be imposed on the brane, see
e.g.~\cite{DerMad03} for a review.

In this paper we describe gravity in the five dimensional bulk by means of the
Einstein Gauss-Bonnet equations and use the junction conditions obtained
in~\cite{Dav02} (which are briefly reviewed in section 2).
To obtain the geometry of an almost flat brane imbedded in a perturbed
anti-de Sitter bulk and, hence, its gravitational field, we use the
approach advocated in~\cite{DerDoz01}. 

More precisely, we first construct in section 3 the most general linear
perturbations of the bulk satisfying the Einstein Gauss-Bonnet equations~; it
turns out, as already known,\cite{KimLee00,Neu01} that they are the
same as in Einstein's theory (the bulk background being locally anti-de Sitter).
Then, in section 4, we use the junction
conditions obtained in \cite{Dav02}
 and write the equations which govern gravity on
the brane~; we deduce from them a set of equations that can be compared to the
usual four dimensional linearized Einstein equations~; we see, as already
known,\cite{KimLee00,Neu01} that the two sets (after proper
identification of Newton's constant)are identical if only
``zero modes" are present in the bulk (however only very special matter
sources on the brane are compatible with zero modes only).

We then turn, in section 5, to the case when matter on the brane is a static,
spherically symmetric, point source and give the linearized gravitational
potentials in terms of well defined integrals which generalize the results
obtained in~\cite{DerDoz01} (equations (\ref{4metric}-\ref{cUV}) in text).
 We finally present
our main results in sections 6 and 7~: by approximating
the integrands by simpler functions we are able to perform the integrations and
thus obtain the gravitational potentials in analytic form valid for {\sl all}
distances.

This allows us to  see that, when  Einstein's theory governs the bulk, the
``correction" to Newton's potential on the brane goes
 from its large $1/r$ behaviour at small $r$ to its
$1/r^2$ one at large $r$  quite slowly~; in the Einstein Gauss-Bonnet case
on the other hand, we find that the correction to Newton's potential
 can be small for {\sl all} $r$.
We hence conclude that the  Einstein ``Lanczos Lovelock Gauss-Bonnet" equations
in the bulk (rather than simply Einstein's) induce on the brane a better
approximation to Newton's law and the usual linearized {\it four}-dimensional
Einstein equations.

Throughout this paper, we denote double partial derivatives
$\partial_\mu\partial_\nu$ by $\partial_{\mu\nu}$.

\section{The Einstein ``Lanczos Lovelock Gauss-Bonnet" equations and
 the associated junction conditions}

To construct a ``$Z_2$-symmetric braneworld" one can proceed as follows~: 
consider a five dimensional Lorentzian manifold $V_+$ with a timelike edge, and  
superpose $V_+$
and its copy $V_-$ onto each other along the edge~; one thus obtains a braneworld, 
composed of a $Z_2$-symmetric spacetime or ``bulk" $V_5$, and a singular surface,
or ``brane" $\Sigma_4$ whose extrinsic curvature is discontinuous~: the extrinsic
curvature of $\Sigma_4$ embedded in $V_-$ is the opposite of the extrinsic curvature
of $\Sigma_4$ embedded in $V_+$. Some components of the braneworld Riemann tensor 
therefore exhibit a delta-type singularity in $V_5\ {\cal t}\ \Sigma_4$. In a Gaussian
normal coordinate system such that the equation defining the position of the brane is
$y=0$ the braneworld line element can be written as
\begin{eqnarray}
ds^2|_5=dy^2+\bar g_{\mu\nu}(x^\rho, |y|)dx^\mu dx^\nu
\label{gnmetric}
\end{eqnarray}
where $y>0$ spans $V_+$ and $y<0$ spans $V_-$. As for the brane extrinsic curvatures
 in $V_+$ and $V_-$, they are given by
\begin{eqnarray}
 K_{\mu\nu}\equiv K^+_{\mu\nu}
=-K^-_{\mu\nu}=-{1\over2}{\partial\bar g_{\mu\nu}\over\partial y}|_{{y=0}_+}\,.
\label{extcurv}
\end{eqnarray}

We shall associate to this braneworld the following action
\begin{eqnarray}
S=\int_{V_5}\!d^5x\sqrt{-g}\,L_{[2]}+2\kappa\!\int_{\Sigma_4}\!d^4x
\sqrt{-\bar g}\,{\cal L}_m-2\int_{\Sigma_4}\!d^4x\sqrt{-\bar g}\,Q\,.
\end{eqnarray}
$g$ is the determinant of the  bulk metric coefficients $g_{AB}$, $\bar g$ that
of the induced brane  metric coefficients $\bar g_{\mu\nu}$. In the first term~:
\begin{eqnarray}
&&L_{[2]}\equiv-2\Lambda+s+\alpha L_{(2)}\,,
\cr
&&\quad\hbox{where}\quad L_{(2)}\equiv s^2-4r^{LM}r_{LM}+R^{LMNP}R_{LMNP}\,,
\end{eqnarray}
is the Einstein Lanczos Lagrangian (also called Einstein Gauss-Bonnet
and generalized by Lovelock~\cite{Lan32}, see e.g.~\cite{DerMad03}
for a review). Here 
$R^A{}_{BCD}\equiv\partial_C\Gamma^A_{BD}-...$ 
are the components of the Riemann tensor, 
$\Gamma^A_{BD}$ being the Christoffel symbols and all indices $A$ being
 moved with $g_{AB}$ and its inverse $g^{AB}$~;
$r_{BD}\equiv R^A{}_{BAD}$ are the Ricci tensor components, 
$s\equiv g^{BD}r_{BD}$ is the scalar curvature~; $\Lambda$ is the bulk 
``cosmological constant" and $\alpha$ a $(length)^2$ parameter
 (that we shall take to be positive, see e.g.~\cite{Zwi85}).
In the second term, ${\cal L}_m(\bar g_{\mu\nu})$ is
the brane ``tension plus matter" Lagrangian and $\kappa$  a coupling constant
which will be ultimately related to Newton's. The third, boundary, term,
which generalizes Gibbons-Hawking's~\cite{GibHaw77}, was first obtained
by Myers~\cite{Mye87} and its explicit expression is (see~\cite{Dav02})
\begin{eqnarray}
Q\equiv2K+4\alpha(J-2\bar\sigma^\mu_\nu \,K^\nu_\mu)
\end{eqnarray}
 with $J$  the trace of
\begin{eqnarray}
 J^\mu_\nu\equiv -{2\over3}K^\mu_\rho K^\rho_\sigma K^\sigma_\nu
+{2\over3}K\,K^\mu_\rho K^\rho_\nu+{1\over3}K^\mu_\nu(K.K-K^2)\,;
\end{eqnarray}
$\bar\sigma^\mu_\nu\equiv\bar r^\mu_\nu-{1\over2}\delta^\mu_\nu\bar s$
is the intrinsic Einstein tensor of the brane $\Sigma_4$ and $K^\mu_\nu$ its
extrinsic curvature (\ref{extcurv}),
 all indices $\mu$ being moved with $\bar g_{\mu\nu}$
and its inverse $\bar g^{\mu\nu}$.

Thanks to this boundary term the variation of $S$ with respect to the metric
coefficients is given in terms of their variations only~\cite{Mye87},
as~\cite{Dav02}~:
\begin{eqnarray}
\delta S=\int_{V_5}\!d^5x\sqrt{-g}\,\sigma^{[2]}_{AB}\,\delta g^{AB}
+\int_{\Sigma_4}\!d^4x\sqrt{-\bar g}\,(2B_{\mu\nu}-\kappa
T_{\mu\nu})\,\delta\bar g^{\mu\nu}\,.
\end{eqnarray}
The ``braneworld equations of motion" are therefore simply taken to be
$\delta S=0$, with the metric fixed at the boundaries at infinity only.
They are, first, the Einstein Gauss-Bonnet ``bulk" equations
(which are second order and quasi-linear in the metric coefficients,
see e.g.~\cite{DerMad03} for a review)~:
\begin{eqnarray}
\sigma^A_{[2]B}\equiv\Lambda \delta^A_B+\sigma^A_B+\alpha\, \sigma^A_{(2)B}=0\,,
\label{bulkeq}
\end{eqnarray}
where $\sigma^A_B\equiv r^A_B-{1\over2}\delta^A_Bs$ is  the bulk Einstein
tensor and
\begin{eqnarray}
\sigma^A_{(2)B}
\equiv 2\left[R^{ALMN}R_{BLMN}-2r^{LM}R^A_{\ LBM}-2 r^{AL}r_{BL}+
sr^A_B\right]-{1\over2}\delta^A_B L_{(2)}
\end{eqnarray}
is the Lanczos tensor. As for the brane equations, which, as shown
 in~\cite{Dav02}, generalize the Israel junction conditions~\cite{Isr66}
 to Einstein Gauss-Bonnet gravity, they are~:
\begin{eqnarray}
B^\mu_\nu\equiv K^\mu_\nu-K\delta^\mu_\nu
+4\alpha\left({3\over2}J^\mu_\nu-{1\over2}J\delta^\mu_\nu-
\bar P^\mu_{\rho\nu\sigma}K^{\rho\sigma}\right)={\kappa\over2}T^\mu_\nu
\label{genjunc}
\end{eqnarray}
 where
\begin{eqnarray}
\bar P_{\mu \rho \nu \sigma }
&\equiv&
 \bar R_{\mu \rho \nu \sigma } +(\bar r_{\mu \sigma }\bar g_{\rho\nu}
-\bar r_{\rho \sigma }\bar g_{\mu \nu }+\bar r_{\rho\nu}
\bar g_{\mu\sigma }-\bar r_{\mu\nu}\bar g_{\rho\sigma })
\cr
&&\quad
-{1\over2}\bar s\,(\bar g_{\mu \sigma }\bar g_{\rho\nu}
-\bar g_{\mu \nu}\bar g_{\rho\sigma})
\end{eqnarray}
 and  where $T_{\mu\nu}$ is defined by
 $\delta(\sqrt{-\bar g}{\cal L}_m)
\equiv-{1\over2}\sqrt{-\bar g}\,T_{\mu\nu}\,\delta\bar g^{\mu\nu}$
 and is interpreted as the stress-energy tensor of ``tension plus matter"
 on the brane. By making a ($4+1$) decomposition of $\sigma^A_{[2]B}$
in the Gaussian normal coordinate system (\ref{gnmetric}),
one can see that 
\begin{eqnarray}
\sigma^\mu_{[2]w}=-\nabla_\nu B^{\mu\nu}
\end{eqnarray}
 with
$\nabla_\nu$ the covariant derivative associated with the brane metric
 $\bar g_{\mu\nu}$, so that the conservation of $T_{\mu\nu}$,  
\begin{eqnarray}
\nabla_\nu T^{\mu\nu}=0\,,
\end{eqnarray}
is included in the bulk equations (\ref{bulkeq}).

If now the bulk is locally an anti-de Sitter (AdS$_5$) spacetime, then,
 because of maximal symmetry~:
\begin{eqnarray}
R_{ABCD}=-{1\over {\cal L}^2}(g_{AC}g_{BD}-g_{AD}g_{BC})\,,
\end{eqnarray}
with
\begin{eqnarray}
{1\over {\cal L}^2}={1\over4\alpha}\left(1\pm\sqrt{1+{4\alpha\Lambda\over3}}\right)
\label{calL}
\end{eqnarray}
in order to satisfy the bulk equations (\ref{bulkeq}).
 (One usually chooses the lower sign so that
 $\lim_{\alpha\to0}{\cal L}^2=-{6\over\Lambda}$, that is the
Einsteinian value~; however, there is no  reason at that level to reject 
the upper sign which allows, in particular,
for an AdS$_5$ bulk even when $\Lambda=0$~; as for the case ${\cal
 L}^2=4\alpha$ it has been analyzed in e.g.~\cite{Zan00}.) 
If, moreover, one wants the brane to be flat, a convenient
 coordinate system to describe the anti-de Sitter bulk is 
\begin{eqnarray}
ds^2|_5=dy^2+e^{-2|y|/{\cal L}}\eta_{\mu\nu}\, dx^\mu dx^\nu
\end{eqnarray}
in which the brane is located at $y=0$ and has an extrinsic curvature given
 $K^\mu_\nu={1\over{\cal L}}\delta^\mu_\nu$, see (\ref{extcurv}). The brane 
equations (\ref{genjunc}) then tell us that 
\begin{eqnarray}
\kappa T^\mu_\nu=-{6\over{\cal L}}\delta^\mu_\nu
\left(1-{4\alpha\over3{\cal L}^2}\right)
\end{eqnarray}
which is interpreted as the ``tension" required for keeping a brane flat
 in an anti de-Sitter bulk governed by the Einstein Gauss-Bonnet equations.

\section{The bulk perturbations}
\medskip

In (quasi) conformally Minkowskian coordinates $x^A=\{x^\mu=(t, x^i), x^4=w\}$,
 the line element of our perturbed
$Z_2$-symmetric anti-de Sitter bulk can be taken to read
\begin{eqnarray}
ds^2|_5= \left({{\cal L}\over w}\right)^2(\eta_{AB}+\gamma_{AB})\,dx^Adx^B
\qquad\hbox{with}\qquad\gamma_{ww}=\gamma_{w\mu}=0
\label{bulkmetric}
\end{eqnarray}
 where ${\cal L}(>0)$  is given by (\ref{calL}) and  where $\gamma_{\mu\nu}$ 
are ten fonctions of the five coordinates $(x^\mu,w)$. Since the (unperturbed)
 brane is located at $w=0$, the coefficients $\gamma_{\mu\nu}$ will be at
first order  functions of $|w|$ only.

If we now impose this metric to satisfy the bulk Einstein Gauss-Bonnet
 equations (\ref{bulkeq}),
 it is an easy exercise to see, first, that the conditions 
(all indices being from now on raised with $\eta^{\mu\nu}$) 
\begin{eqnarray}
\gamma^\rho_\rho=0\qquad,\qquad\partial_\rho\gamma^\rho_\mu=0
\label{bgauge}
\end{eqnarray}
solve the constraints $\sigma^w_{[2]w}=0, \sigma^w_{[2]\mu}=0$ at linear
order and reduce the ten metric coefficients
$\gamma_{\mu\nu}$ to five, which represent the five degrees of freedom of
AdS$_5$ gravitational waves in Einstein Gauss-Bonnet theory (this being so because
the Lanczos-Lovelock Lagrangian yields second order field equations,
just like Einstein-Hilbert's).

As for their  evolution equation, it then reads (with
$\square_4\equiv\eta^{\mu\nu}\partial_\mu\partial_\nu$)
\begin{eqnarray}
\left(1-{4\alpha\over{\cal L}^2}\right)
\left(\square_4\gamma_{\mu\nu}
+\partial_{ww}\gamma_{\mu\nu}-{3\over w}\partial_w\gamma_{\mu\nu}\right)=0\,.
\label{bulkpeq}
\end{eqnarray}
 The presence of the overall coefficient
 $(1-4\alpha/{\cal L}^2)=\mp\sqrt{1+\alpha\Lambda/3}$ is a
consequence of the quasi-linearity of the Lanczos-Lovelock
 equations (\ref{bulkeq}),
 see e.g~\cite{DerMad03}. The case when the parameters $\alpha$ and $\Lambda$
 are such that it vanishes (studied in e.g.~\cite{Zan00}) will not be
 considered here. Therefore, the evolution equation for the 
 AdS$_5$ gravitational waves in Einstein and Einstein Gauss-Bonnet theory are the
same~\cite{KimLee00,Neu01}.

As emphasized in~\cite{DerDoz01}, the general solution of (\ref{bulkpeq})
 must converge (more precisely be $L_2$) on its domain of definition.
 Now, the bulk is composed
 of $V_+$ and
its copy $V_-$ joined by the brane $\Sigma_4$ and its $Z_2-$symmetric metric
 (\ref{bulkmetric})
 depends only on $|w|$. The general solution of (\ref{bulkpeq}) 
is therefore the union of two sets.
First, the set~:
\begin{eqnarray}
\gamma_{\mu\nu}(x^\mu,w)
&=&{\cal R}e
\int\! {d^3k\over(2\pi)^{3\over2}}\,
{dm\over(2\pi)^{1\over2}}\,{\rm e}^{{\rm i}k_\rho x^\rho}e_{\mu\nu}\,w^2Z_2(m|w|)
\label{gamform}
\\
\mbox{with}&&\quad k_0=-\sqrt{k^2+m^2}\quad,\quad k^\rho e_{\rho\mu}=0
\quad,\quad\eta^{\rho\sigma}e_{\rho\sigma}=0
\nonumber
\end{eqnarray}
where $k^2\equiv k_ik^i$, where the five polarisations $e_{\mu\nu}$ 
are a priori arbitrary functions of
 $k^i$ and $m$ and where $Z_2(mw)=H^{(1)}_2(mw)+a_mH^{(2)}_2(mw)$ is an a priori
 arbitrary linear combination of second order Hankel functions of first
and second kinds~\cite{GraRyz80}. The coefficient $a_m$ is determined by
 the model at hand~ and  one usually eliminates
the mode coming from $+\infty$, that is sets $a_m=0\,$. The ``zero modes"
 do not depend on $w$ and are~:
\begin{eqnarray}
&&\gamma^{(0)}_{\mu\nu}(x^\mu)={\cal R}e
\int\! {d^3k\over(2\pi)^{3\over2}}\,{\rm e}^{{\rm i}(k_i x^i-kt)}e_{\mu\nu}(k^i)
\cr
&&\quad
\hbox{with}\quad ke_{0\mu}+k^ie_{i\mu}=0\ ,\ \eta^{\rho\sigma}e_{\rho\sigma}=0\,.
\end{eqnarray}

The second set is composed of the extra $L_2$ modes which converge
exponentially as $|w|\to+\infty$~:
\begin{eqnarray}
\gamma_{\mu\nu}(x^\mu,w)
&=&{\cal R}e\!\int\! {d^3k\over(2\pi)^{3\over2}}
\int_{-k^2}^0\!{dA\over(2\pi)^{1\over2}}\,{\rm e}^{{\rm i}k_\rho
x^\rho}\!e_{\mu\nu}\,w^2H^{(1)}_2({\rm i}\sqrt{|A|}\,|w|)
\\
\mbox{with}&&\quad k_0=-\sqrt{k^2+A}\quad,\quad k^\rho e_{\rho\mu}=0
\quad,\quad\eta^{\rho\sigma}e_{\rho\sigma}=0\,.
\label{gamsol}\nonumber\end{eqnarray}
The static modes do not depend on time and are~:
\begin{eqnarray}
\gamma^{(s)}_{\mu\nu}(x^i,w)={\cal R}e\int\!
 {d^3k\over(2\pi)^{3\over2}}{\rm e}^{{\rm i}k_i x^i}
e_{\mu\nu}(k^i)\,w^2H^{(1)}_2({\rm i}k|w|)\,.
\label{gstatic}
\end{eqnarray}

\section{The linearized equations for gravity on an Einstein Gauss-Bonnet brane}

 The position of the brane $\Sigma_4$ in the bulk $V_5$ is defined in the 
coordinate system (\ref{bulkmetric}) by
\begin{eqnarray}
w={\cal L}+\zeta(x^\mu)
\end{eqnarray}
where the function $\zeta(x^\mu)$ is a priori arbitrary and describes the
 so-called ``brane-bending" effect.
The induced metric on $\Sigma_4$ is, at linear order
\begin{eqnarray}
ds^2|_4=(\eta_{\mu\nu}+h_{\mu\nu})dx^\mu dx^\nu\qquad\hbox{with}\qquad h_{\mu\nu}
=\gamma_{\mu\nu}|_{\Sigma} -
2{\zeta\over{\cal L}}\eta_{\mu\nu}
\label{hmetric}
\end{eqnarray}
where $\gamma_{\mu\nu}(x^\mu,w)$ is a solution of
 (\ref{bgauge}-\ref{bulkpeq})
 and where the index
 $\Sigma$ means that the quantity is evaluated at $w={\cal L}$.
 As for its extrinsic curvature
(defined by (\ref{extcurv}) in a Gaussian normal coordinate system) it is given,
 in the system (\ref{bulkmetric}-\ref{bgauge}), by~\cite{GS01}
\begin{eqnarray}
K^\mu_\nu={1\over{\cal L}}\delta^\mu_\nu
-{1\over2}\partial_w\gamma^\mu_\nu|_\Sigma+\partial^\mu_\nu\zeta\,.
\label{Kpert}
\end{eqnarray}

Now, as we have seen in section 2, the ``brane equations" are (\ref{genjunc}). 
In the case at hand $K^\mu_\nu$ is given by (\ref{Kpert}) and 
$\bar P^\mu_{\ \rho\nu\sigma}$ is to be computed by
means of the metric (\ref{hmetric}). Splitting $T^\mu_\nu$ into a brane 
 tension and matter stress-energy tensor $S^\mu_\nu$ as
\begin{eqnarray}
\kappa T^\mu_\nu\equiv -{6\over{\cal L}}\delta^\mu_\nu
\left(1-{\bar\alpha\over3}\right)+\kappa S^\mu_\nu
\end{eqnarray}
where, from now on we shall use the notation
\begin{eqnarray}
\quad\bar\alpha\equiv{4\alpha\over{\cal L}^2}\,,
\end{eqnarray}
the brane equations (\ref{genjunc}) read, at linear order
\begin{eqnarray}
\kappa S^\mu_\nu=2\left(1+{\bar\alpha}\right)
(\partial^\mu_\nu\zeta-\delta^\mu_\nu\square_4\zeta)
-\left(1-{\bar\alpha}\right)\partial_w
\gamma^\mu_\nu|_\Sigma-\bar\alpha{\cal L}\,\square_4\gamma^\mu_\nu|_\Sigma\,.
\label{Smn}
\end{eqnarray}

Equations (\ref{hmetric}) and (\ref{Smn})
  together with (\ref{gamform}) and (\ref{gamsol})
completely
 describe gravity on the brane. They have two useful consequences
\begin{eqnarray}
\partial_\rho S^\rho_\nu=0\qquad,\qquad \kappa S
=-6\left(1+{\bar\alpha}\right)\square_4\zeta\,.
\label{Slaw}
\end{eqnarray}

Let us compare them  to the usual four dimensional linearized Einstein equations.
 In order to do so, we first perform a coordinate transformation
  $x^\mu\to x^\mu=x^{*\mu}+\epsilon^\mu$ such that the new metric coefficients
\begin{eqnarray}
h^*_{\mu\nu}=\gamma_{\mu\nu}|_\Sigma-{2\over{\cal L}}\zeta\eta_{\mu\nu}
+\partial_\mu\epsilon_\nu+\partial_\nu\epsilon_\mu\qquad\hbox{with}
\qquad\square_4\epsilon_\mu=-{2\over{\cal L}}\partial_\mu\zeta
\label{hstar}
\end{eqnarray}
 satisfy the harmonicity condition
\begin{eqnarray}
\partial_\mu \left(h^{*\mu}_\nu-{1\over2}\delta^\mu_\nu h^*\right)=0\,.
\end{eqnarray}
Taking the d'Alembertian of (\ref{hstar}), eliminating $\zeta$ by means of
 (\ref{Smn}) and using (\ref{bgauge}-\ref{bulkpeq})
  we get the following consequence of the brane equations 
\begin{eqnarray}
\square_4h^*_{\mu\nu}
&=&-\left({2\over1+\bar\alpha}\right){\kappa\over{\cal
 L}}\left(S_{\mu\nu}-{1\over2}\eta_{\mu\nu}S\right)
\cr
&&\quad+\left({1-\bar\alpha\over1+\bar\alpha}\right)\left[
-(\partial_{ww}\gamma_{\mu\nu})|_\Sigma
+{1\over{\cal L}}(\partial_w\gamma_{\mu\nu})|_\Sigma\right]
\,.\label{effeq}
\end{eqnarray}

Now, recall that the usual linearized Einstein equations on a four dimensional
Minkowski background read, in harmonic coordinates
\begin{eqnarray}
\square_4h^*_{\mu\nu}=-16\pi G\left(S_{\mu\nu}-{1\over2}\eta_{\mu\nu}S\right)
\label{linEin}
\end{eqnarray}
where $G$ is Newton's constant. One therefore sees that, if 
we identify, in agreement with~\cite{Neu01}
\begin{eqnarray}
8\pi G\equiv\left({1\over1+\bar\alpha}\right){\kappa\over{\cal L}}
\label{8piG}\,,
\end{eqnarray}
and if only zero modes are present in the bulk,
 then (\ref{effeq}) reduces to (\ref{linEin}).

There is however an important proviso. 
The usual Einstein equations (\ref{linEin})
 hold for {\it any} type of matter (compatible with the harmonicity condition,
 or, equivalently with the Bianchi identity $\partial_\rho S^\rho_\mu=0$. 
 By contrast, the linearized equations for gravity on a brane are (\ref{effeq}),
 {\it provided} the source $S_{\mu\nu}$ satisfies the junction
 condition (\ref{Smn}).
 If only zero modes are present in the bulk, then, since the last two
 terms in (\ref{Smn}) are absent because of (\ref{bulkpeq}), the derivatives
 $\partial_\lambda({\cal S}_{\mu\nu}-{1\over3}\eta_{\mu\nu} {\cal S})$ must be
symmetric in $\lambda$ and $\mu$, a property which is satisfied only by very
 special matter, obeying a very contrived equation of state
 (see e.g.~\cite{DerDoz01} for details). There is therefore no
compelling reason, at that level, to make the identification (4.12),
that we shall consider until further notice as a mere notation.

Now, of course, the junction conditions are better seen as boundary conditions
 on the allowed modes in the bulk. Indeed, if matter on the brane is known,
 then (\ref{Smn}-\ref{Slaw}) 
 can be inverted to give $\zeta$ and the last two terms in (\ref{Smn})
in terms of $S_{\mu\nu}$. With that information one can in principle get,
 by inversion of (\ref{gamform}) and (\ref{gamsol}), the
polarisations $e_{\mu\nu}$, and hence the allowed bulk perturbations, in terms of
 the brane matter source. Then the induced metric on the brane (\ref{hmetric})
 is known in
 terms of the matter variables and can be compared with the usual
 four dimensional Einstein result.
 This programme is completed below in the particular case of a point static source.

\section{The gravitational potentials of
 a static point source on an Einstein Gauss-Bonnet brane}

Consider a static, spherically symmetric, point-like source on the brane.
 Its stress-energy tensor is, (with $\vec r=\{x^i\}$, $r^2=x_ix^i$)~:
\begin{eqnarray}
S_{00}=M\delta(\vec r)\qquad,\qquad S_{0i}=S_{ij}=0\,.
\end{eqnarray}
When necessary, we shall go to Fourier space~:
\begin{eqnarray*}
&&f\leftrightarrow\hat f\quad
\mbox{with}\quad
 f=\int\!{d^3k\over{(2\pi)^{3/2}}}{\rm e}^{{\rm i}k_ix^i}\hat f\,, 
\cr
&&\quad
\hat\delta={1\over(2\pi)^{3/2}}\,,
\quad
\widehat{1\over r}=\sqrt{2\over\pi}{1\over k^2}
\quad
\mbox{and}
\quad
\widehat{\partial_{ij}{1\over r}}=-\sqrt{2\over\pi}{k_ik_j\over k^2}\,.
\end{eqnarray*}

The junction conditions (\ref{Smn}-\ref{Slaw}) then give~:
\begin{eqnarray}
&&\zeta=-{1\over24\pi}{\kappa M\over(1+\bar\alpha)}{1\over r}
\quad,
\cr
&&\left(1-{\bar\alpha}\right)\partial_w\gamma^{(s)}_{ij}|_\Sigma
+\bar\alpha{\cal L}
\triangle\gamma^{(s)}_{ij}|_\Sigma
=-{\kappa M\over 3}\delta (\vec r)\delta_{ij}
-{\kappa M\over12\pi}\partial_{ij}{1\over r}
\cr
\cr
&&\left(1-{\bar\alpha}\right)\partial_w\gamma^{(s)}_{00}|_\Sigma
+\bar\alpha{\cal L}\triangle\gamma^{(s)}_{00}|_\Sigma
=-{2\kappa M\over 3}\delta
(\vec r)\quad,
\cr
&&\left(1-{\bar\alpha}\right)\partial_w\gamma^{(s)}_{0i}|_\Sigma
+\bar\alpha{\cal L}\triangle\gamma^{(s)}_{0i}|_\Sigma=0\,.
\label{gameq}
\end{eqnarray}

Now, the static bulk modes are given by (\ref{gstatic})~: 
$\hat\gamma^{(s)}_{\mu\nu}(k^i,w)=e_{\mu\nu}(k^i)\,w^2H_2^{(1)}({\rm i}k|w|)$
 and their $w$-derivatives on $\Sigma$
are~\cite{GraRyz80}~:
$\partial_w\hat\gamma^{(s)}_{\mu\nu}|_\Sigma
=e_{\mu\nu}(k^i)\,{\rm i}k{\cal L}^2\,H_1^{(1)}({\rm i}k{\cal L})$.
 Equation (\ref{gameq}) therefore gives the polarisations in terms of the
brane stress-energy tensor as
\begin{eqnarray}
e_{00}(k^i)\left[(1-\bar\alpha)\,H_1^{(1)}({\rm i}k{\cal L})
+{\rm i}\,\bar\alpha\,k{\cal L}
H_2^{(1)}({\rm i}k{\cal L})\right]
&=&-{2\over 3}{\kappa M\over(2\pi)^{3\over2}}{1\over {\rm i}k{\cal L}^2}
\quad,\qquad e_{0i}(k^i)=0
\cr\cr
e_{ij}(k^i)\left[(1-\bar\alpha)
\,H_1^{(1)}({\rm i}k{\cal L})+{\rm i}\,\bar\alpha\,
k{\cal L}H_2^{(1)}({\rm i}k{\cal L})\right]&=&-{1\over 3}
{\kappa M\over(2\pi)^{3\over2}} {1\over {\rm i}k{\cal L}^2}
\left(\delta_{ij}-{k_ik_j\over k^2}\right)\,.
\end{eqnarray}
 The bulk metric  is then known and is~:
\begin{eqnarray}
&&\gamma_{\mu\nu}(x^i,w)={\cal R}e\int\!{d^3k\over(2\pi)^{3\over2}}
\,e^{{\rm i}k_ix^i}\hat\gamma_{\mu\nu}(k^i, w)
\cr
\cr
\hbox{with}\qquad
&&\hat\gamma_{\mu\nu}(k^i,w)=
{\kappa M\over3{\cal L}(2\pi)^{3\over2}}\,{w^2\,K_2(k|w|)\over k{\cal L}
\left[(1-\bar\alpha)K_1(k{\cal L})+\bar\alpha\,k{\cal L}K_2(k{\cal L})\right]}
\,c_{\mu\nu}
\end{eqnarray}
 where $c_{00}=2$, $c_{0i}=0$ and $c_{ij}=\delta_{ij}-k_ik_j/k^2$
 and where $K_\nu(z)$ is the modified Bessel function defined as
$K_\nu(z)={\rm i}{\pi\over2}e^{{\rm i}\nu{\pi\over2}}H_\nu^{(1)}({\rm i}z)$. 

The $\hat h_{00}$ Fourier component of the metric on the brane therefore reads
\begin{eqnarray}
\hat h_{00}(\vec k)
&=&\hat\gamma_{00}|_\Sigma+2{\hat\zeta\over{\cal L}}
\cr
&=&{\kappa M\over k^2{\cal L}(2\pi)^{3\over2}}{1\over(1+\bar\alpha)}
\left[1+{2\over3}
{(1-\bar\alpha)\,k{\cal L}\,K_0(k{\cal L})\over (1+\bar\alpha)K_1(k{\cal
L})+\bar\alpha\,k{\cal L}\,K_0(k{\cal L})}\right]\,
\end{eqnarray}
and a similar expression for $\hat h_{ij}$.

We then take the Fourier transform and integrate over angles (after
 elimination of the $k_ik_j$ term in $\hat h_{ij}$ by a suitable coordinate
 transformation). We then go to
Schwarzschild coordinates by means of yet another infinitesimal coordinate
 transformation. Setting $x=r/{\cal L}$ and recalling that we
 introduced the notation
$\left({1\over1+\bar\alpha}\right){\kappa\over{\cal L}}\equiv 8\pi G$, we finally
 obtain the linearized metric on the brane, as created by a static point-like
 source, as~:
\begin{eqnarray}
ds^2|_4=-(1+2U)dt^2+(1-2V)dr^2+r^2(d\theta^2+\sin^2\theta d\phi^2)
\label{4metric}
\end{eqnarray}
with
\begin{eqnarray}
U=-{GM\over r}\left[1+{4\over3\pi}{\cal U}_{\bar\alpha}(x)\right]
\quad,\quad
 V=-{GM\over r}\left[1+{2\over3\pi}{\cal V}_{\bar\alpha}(x)\right]
\label{UandV}
\end{eqnarray}
and
\begin{eqnarray}
{\cal U}_{\bar\alpha}(x)
&=&(1-\bar\alpha)\int_0^{+\infty}\!du\,\sin(ux)
\,{K_0(u)\over
(1+\bar\alpha)K_1(u)+\bar\alpha\,u\,K_0(u)}
\cr
{\cal V}_{\bar\alpha}(x)
&=&{\cal U}_{\bar\alpha}(x)-x\,\partial_x\,{\cal U}_{\bar\alpha}(x)\,.
\label{cUV}
\end{eqnarray}
${\cal U}_{\bar\alpha}(x)$ is a well defined integral that we shall
 evaluate analytically.

\section{The  corrections to Newton's law on an Einstein brane for all distances}

When the bulk obeys the five dimensional Einstein equations, that is when
 $\bar\alpha=0$, ${\cal U}_{\bar\alpha}(x)$ in (\ref{cUV}) reduces
 to~\cite{DerDoz01}
\begin{eqnarray}
{\cal U}_0(x)=\lim_{\epsilon\to0}\int_0^{+\infty}\!du\,\sin(ux)
\,U_0(u)\,e^{-\epsilon u}\quad\hbox{with}\quad U_0(u)
\equiv{K_0(u)\over K_1(u)}\,.
\label{calU0}
\end{eqnarray}

By approximating $U_0(u)$ by its asymptotic expressions, that is 
$U_0(u)\sim 1$ when $u\to\infty$ and $U_0(u)\sim-u\ln u$ when $u\to0$,
 it is a  straightforward
exercise~\cite{GraRyz80} to see that $\lim_{x\to0}{\cal U}_0(x)=x^{-1}$
 and that $\lim_{x\to\infty}{\cal U}_0(x)=\pi/2x^2$. Hence we recover the
 well-known result~\cite{GarTan99} (see also~\cite{DerDoz01}) that,
 at short distances the ``correction" to the gravitational potentials
 diverges in ${\cal L}/r$
\begin{eqnarray}
\lim_{r/{\cal L}\to0}U=\lim_{r/{\cal L}\to0} V
=-{4\over3\pi}{GM{\cal L}\over r^2}
\label{UV0}
\end{eqnarray}
whereas at distances large compared with the characteristic scale ${\cal L}$ of
 the anti-de Sitter bulk the correction is reduced by another ${\cal L}/r$
factor~\cite{RanSun99b}, so that~\cite{GarTan99}
\begin{eqnarray}
\lim_{r/{\cal L}\to\infty}U=-{GM\over r}\left[1+{2\over3}
\left({{\cal L}\over r}\right)^2\right]\quad,\quad\lim_{r/{\cal L}\to\infty}V
=-{GM\over
r}\left[1+\left({{\cal L}\over r}\right)^2\right]\,.
\label{Elimit}
\end{eqnarray}

However one can do better than just obtaining the asymptotic behaviours of
the potentials. Indeed it turns out, see Fig.~1, that we can use  for all $u$,
to a good approximation
\begin{eqnarray}
U_0(u)\approx\bar U_0(u)\qquad\hbox{with}\qquad\bar U_0(u)
\equiv u\ln\left(1+{1\over u}\right)\,.
\label{U0ap}
\end{eqnarray}
Using that approximation, (\ref{calU0}) can be integrated for all
 $x$ to yield~:
\begin{eqnarray}
{\cal U}_0(x)\approx\frac{x\,\cos x - \sin x}{x^2}\,{\rm ci}(x)
+\frac{\cos x + x\,\sin x}{x^2}\,{\rm si}(x)
+\frac{\pi }{2\,x^2}\,
\label{cU0}
\end{eqnarray}
and
\begin{eqnarray}
{\cal V}_0(x)
&\approx&
 \frac{3x \cos x+ (x^2 - 3)\sin x}{x^2}\, {\rm ci}(x)
\cr
&&\qquad\quad
+\frac{3x \sin x- (x^2 - 3)\cos x}{x^2}\,{\rm si}(x)
 + \frac{3\pi}{2x^2} - \frac{1}{x}\,,
\label{cV0}
\end{eqnarray}
 where ${\rm si}(x)$ and ${\rm ci}(x)$ are the
integral sine and cosine functions, see~\cite{GraRyz80}.

%\vspace{3mm}
\begin{figure}
\begin{center}
\includegraphics[width=12cm]{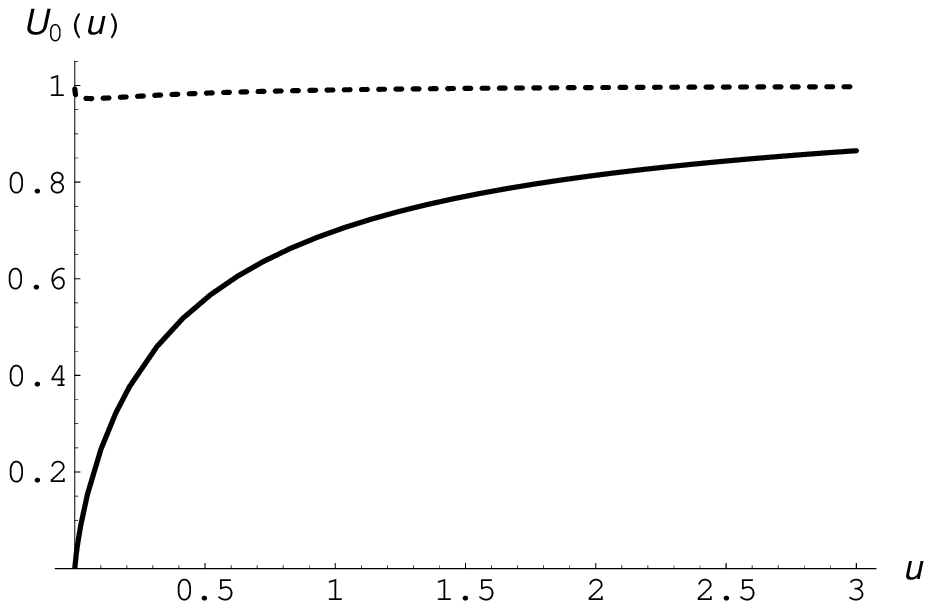}
\\
~\\
\begin{minipage}{13cm}
\hspace*{-5mm}
\small Fig.~1.~The behaviour of $U_0(u)$ and the accuracy of
the approximate analytic formula $\bar{U}_0(u)$. The solid curve is $U_0(u)$
while the upper dotted line shows $\bar{U}_0/U_0$.
One sees that the relative error is small: $\lesssim2\%$ at most.
\end{minipage}
\end{center}
\end{figure}
%\vspace{3mm}

In Fig.~2, ${\cal U}_0(x)$ is plotted as a function of $x=r/{\cal L}$.
 As one can see the transition from the
$1/r$ behaviour at short distances to the $1/r^2$ one at large distances
is quite slow. This is due to the slow convergence of $U_0(u)$ when $u\to0$.
As seen from the upper-most lines of Fig.~4 below, the Newtonian
behaviour is recovered at $r/{\cal L}\gtrsim10$. 

%\vspace{3mm}
\begin{figure}
\begin{center}
\includegraphics[width=12cm]{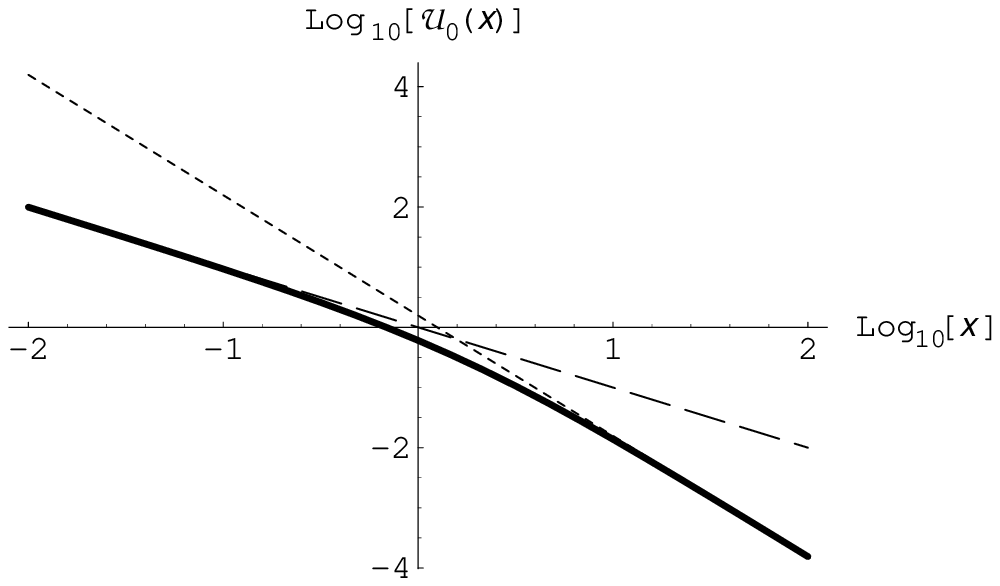}
\\
~\\
\begin{minipage}{13cm}
\hspace*{-5mm}
\small
Fig.~2.~A logarithmic plot of the function ${\cal U}_0(x)$.
The short-dashed curve is $\pi/(2x^2)$ and
the long-dashed curve is $1/x$.
\end{minipage}
\end{center}
\end{figure}

\section{The  corrections to Newton's law on an Einstein Gauss-Bonnet brane}

When the bulk obeys the five dimensional Einstein Gauss-Bonnet equations,
 ${\cal U}_{\bar\alpha}(x)$ in (\ref{cUV}) is
\begin{eqnarray}
&&{\cal U}_{\bar\alpha}(x)
=(1-\bar\alpha)\int_0^{+\infty}\!du\,\sin(ux)\,U_{\bar\alpha}(u)
\cr
&&
\quad\hbox{with}\quad
U_{\bar\alpha}(u)\equiv{K_0(u)
\over \bar\alpha u\,K_0(u)+(1+\bar\alpha)K_1(u)}\,.
\label{cUalp}
\end{eqnarray}

Again, by approximating $U_{\bar\alpha}(u)$  by its asymptotic expressions,
 that is 
\begin{eqnarray*}
U_{\bar\alpha}(u)\sim {1\over\bar\alpha u}
\quad\mbox{when}\quad u\to\infty\,,
\quad\mbox{and}\quad
U_{\bar\alpha}(u)\sim-{1\over1+\bar\alpha}u\ln u
\quad\mbox{when}\quad u\to0,
\end{eqnarray*}
 it is a simple exercise~\cite{GraRyz80} to see that
\begin{eqnarray*}
\lim_{x\to0}{\cal U}_{\bar\alpha}(x)
={\pi\over2}\,{1-\bar\alpha\over\bar\alpha}
\qquad\mbox{and}
\qquad
\lim_{x\to\infty}{\cal U}_{\bar\alpha}(x)
={\pi\over2x^2}{1-\bar\alpha\over1+\bar\alpha}\,.
\end{eqnarray*}
 Hence we recover the known result~\cite{KimLee00,Neu01}
 that at distances large compared with
the characteristic scale
${\cal L}$ of the anti-de Sitter bulk the gravitational potentials are
 corrected by a small $1/r^2$ factor, their precise expressions being
\begin{eqnarray}
\lim_{r/{\cal L}\to\infty}U
&=&-{GM\over r}\left[1+{2\over3}
\left({1-\bar\alpha\over1+\bar\alpha}\right)
\left({{\cal L}\over r}\right)^2\right]\ ,
\cr
\lim_{r/{\cal L}\to\infty}V
&=&-{GM\over r}\left[1+\left({1-\bar\alpha\over1+\bar\alpha}\right)
\left({{\cal L}\over r}\right)^2\right]\,.
\label{llimit}
\end{eqnarray}
When $\bar\alpha=0$, (\ref{llimit}) reduces to its ``Einsteinian'' 
expression (\ref{Elimit}).
On the other hand  the corrections  remain finite when $r/{\cal L}\to0$
 so that the gravitational potentials keep their Newtonian behaviour~: 
\begin{eqnarray}
\lim_{r/{\cal L}\to0}U
&=&-{GM\over r}
\left[1+{2\over3}\left({1-\bar\alpha\over\bar\alpha}\right)\right],
\cr
\lim_{r/{\cal L}\to0}V
&=&-{GM\over r}
\left[1+{1\over3}\left({1-\bar\alpha\over\bar\alpha}\right)\right]
\label{UVshort}
\end{eqnarray}
contrarily to what happens in the Einstein case, see
 equation (\ref{UV0}).

Now, here again, one can do better by approximating $U_{\bar\alpha}(u)$ for all
$u$. For example, for all $0<\bar\alpha<1$ (corresponding to the lower
sign in (\ref{calL})), 
a good approximation (to better than $1\%$) is
\begin{eqnarray}
U_{\bar\alpha}(u)\approx\bar U_{\bar\alpha}(u)\qquad\hbox{with}
\qquad\bar U_{\bar\alpha}(u)
\equiv\left({1\over1+\bar\alpha}\right){u\over1+b\,u}
\ln\left[1+{(1+\bar\alpha)\over\bar\alpha}{b\over u}\right]
\label{Ugap}
\end{eqnarray}
and $b=1-\sqrt{1-\bar\alpha\over1+\bar\alpha}\,.$
 Unfortunately we were not able to integrate (\ref{cUalp})
 analytically with that approximation. We hence used
 another approximate form for $U_{\bar\alpha}(u)$, valid for all
 $\bar\alpha\neq0$ :
\begin{eqnarray}
U_{\bar\alpha}(u)\approx\tilde U_{\bar\alpha}(u)
\qquad\hbox{with}\qquad\tilde U_{\bar\alpha}
\equiv\left({1\over1+\bar\alpha}\right)
\left[u\ln\left(1+{\beta\over u}\right)
 -{\beta u\over u+\beta\gamma}\right]
\label{tildeU}
\end{eqnarray}
where $\beta$ and $\gamma$ are constants such that
\begin{eqnarray*}
&&\beta=\frac{\left(\gamma^2-1/3\right)}{\left(\gamma-1/2\right)^2}\,,
\quad
\left(\gamma^2-\frac{1}{3}\right)^2
-\frac{1+\bar\alpha}{\bar\alpha}\left(\gamma-\frac{1}{2}\right)^3=0
\cr
\quad
&&\left(\hbox{and}\quad\,\beta^2=\frac{1+\bar\alpha}{\bar\alpha}
\frac{1}{\left(\gamma-1/2\right)}\right).
\end{eqnarray*}
The relative error is found to be very small in both ranges
of $u\gg1$ and $u\ll1$ for any $\bar\alpha$.
At mildly small values of $u$, the error becomes
as large as $25\,\%$, but only for very large $\bar\alpha$ (see Fig.~3 below).
Using (\ref{tildeU}), (\ref{cUalp}) becomes
\begin{eqnarray}
{\cal U}_{\bar\alpha}(x)
&\approx&\left(\frac{1-\bar\alpha}{1+\bar\alpha}\right)\,
\Biggl\{\frac{\beta\,x\,\cos (\beta\,x)-\sin(\beta\,x)}{x^2}\,
{\rm ci}(\beta\,x)
+\frac{\beta\,x\,\sin(\beta\,x)+\cos (\beta\,x)}{x^2}\,{\rm si}(\beta\,x)
\cr
&&\qquad
+\frac{\pi}{2x^2}
-\frac{\beta}{x}
+ \beta^2\gamma\,\Bigl[{\rm ci}(\beta\,\gamma\,x)\,
     \sin (\beta\,\gamma\,x) 
-\cos (\beta\,\gamma\,x)\, {\rm si}(\beta\,\gamma\,x)\Bigr]\Biggr\}\,,
\label{cUapprox}
\\
\nonumber\\
{\cal V}_{\bar\alpha}(x)
&\approx&\left(\frac{1-\bar\alpha}{1+\bar\alpha}\right)
\Biggl\{
\frac{
     3\,\beta \,x\,\cos (x\,\beta ) + 
       \left( x^2\,{\beta }^2-3 \right)
          \,\sin (\beta\,x )}{x^2} \,{\rm ci}(\beta\,x )
\cr
&&\hspace{10mm}
+\frac{3\,\beta \,x\,\sin (\beta\,x )-
\left(\beta^2\,x^2-3 \right) \,
        \cos (\beta\,x ) }{x^2} \,{\rm si}(\beta\,x )
+\frac{3\,\pi }{2\,x^2} 
- \frac{3\,\beta }{x}
\cr
&&\hspace{10mm}
+{\beta }^2\,\gamma \,
\Bigl[{\rm ci}(\beta \,\gamma\,x )\,
 \Bigl(\sin (\beta \,\gamma\,x )
       -\beta \,\gamma \,x\,\cos (\beta \,\gamma\,x ) \Bigr)  
\cr
&&\hspace{25mm}-{\rm si}(\beta \,\gamma\,x ) 
   \Bigl( \cos (\beta \,\gamma\,x ) 
+ \beta \,\gamma \,x\,
      \sin (\beta \,\gamma\,x ) \Bigr)\Bigr]\Biggr\}\,.
\label{cVapprox}
\end{eqnarray}

%\vspace{3mm}
\begin{figure}
\begin{center}
\includegraphics[width=12cm]{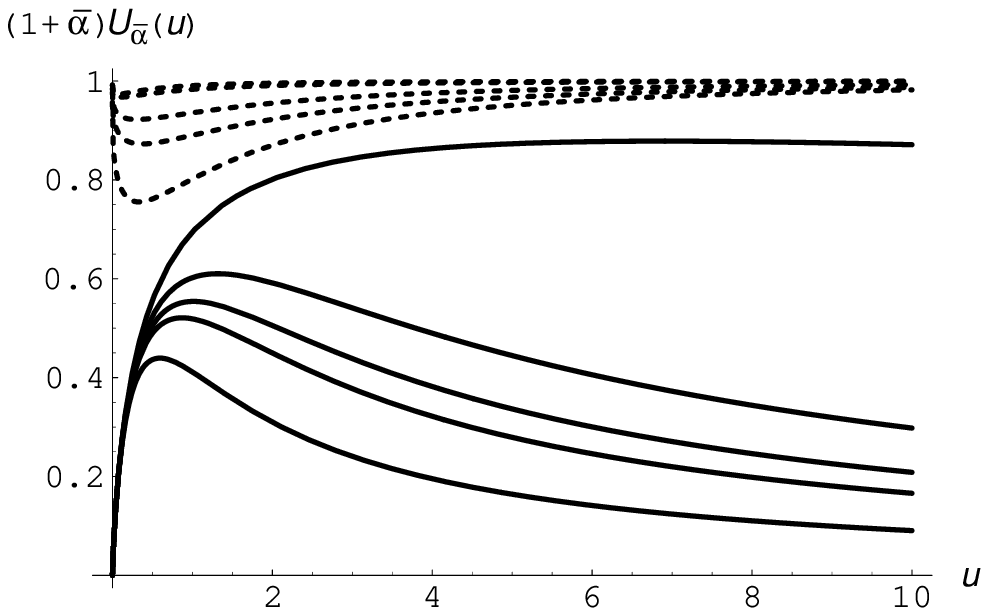}
\\
~\\
\begin{minipage}{13cm}
\hspace*{-5mm}
\small
Fig.~3.~The behaviour of $U_{\bar\alpha}(u)$ for several values
of $\bar\alpha$ and the accuracy of the approximate formula
$\tilde{U}_{\bar\alpha}(u)$.
 The solid lines are
$(1+\bar\alpha)U_{\bar\alpha}(u)$ for the cases 
$\bar\alpha=0.01,\,0.3,\,0.6,\,0.99$ and $\infty$, from top to bottom.
The dotted lines are the ratio $\tilde{U}_{\bar\alpha}/U_{\bar\alpha}$
for the same values of $\bar\alpha$, also from top to bottom.
The upper two dotted lines are almost degenerate in the figure.
\end{minipage}
\end{center}
\end{figure}
%\vspace{3mm}

On the graphs below, see Fig.~4, we plotted $U(r)$ and $V(r)$ 
(as a function of $x=r/{\cal L}$, and divided by the Newtonian
 potential $U_N\equiv-GM/r$). As one can see the gravitational
potentials become closer and closer to their Newtonian values, for all
 distances, when $\bar\alpha\to1$.

%\vspace{3mm}
\begin{figure}
\begin{center}
\includegraphics[width=12cm]{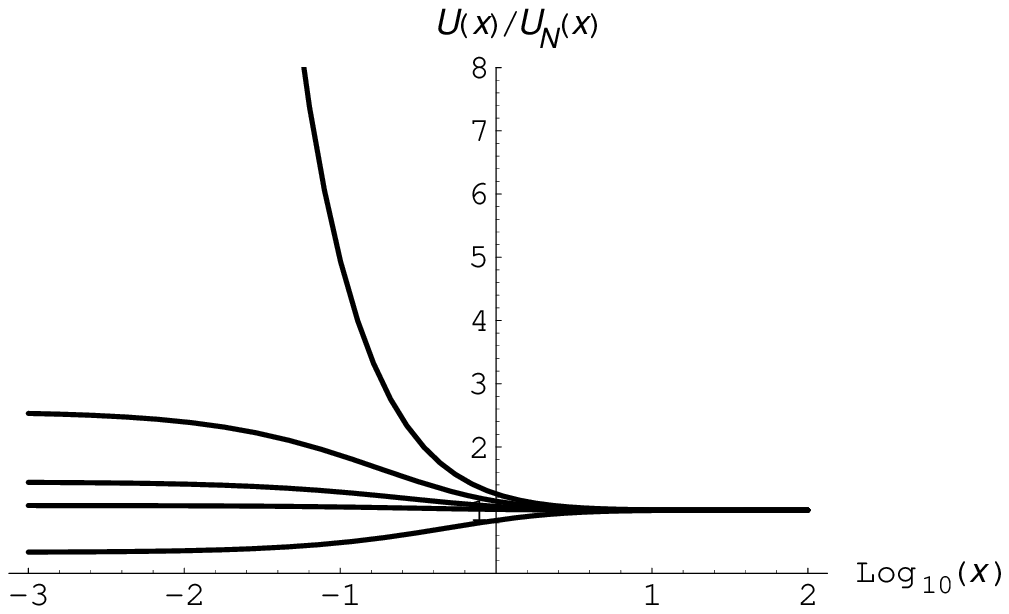}
\\
~\\
\includegraphics[width=12cm]{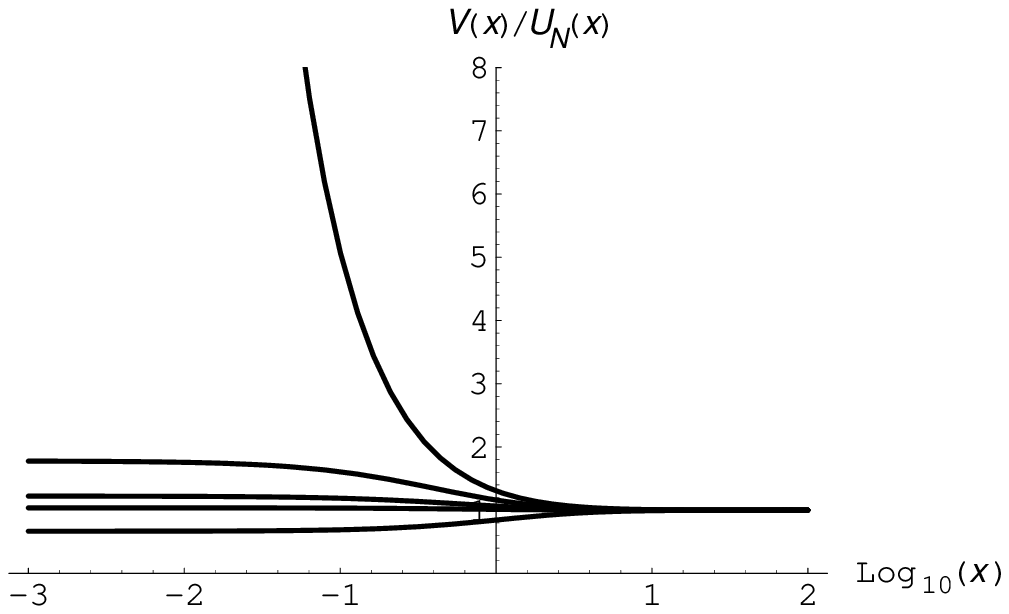}
\\
~\\
\begin{minipage}{13cm}
\hspace*{-5mm}
\small
Figs.~4.~The relative 
corrections to $U$ and $V$ as functions of $x$,
in the upper and lower panels, respectively.
In both figures,
the curves from top to bottom correspond to the cases
$\bar\alpha=0,\,0.3,\,0.6,\,0.9$ and $\infty$.
\end{minipage}
\end{center}
\end{figure}
%\vspace{3mm}

\vspace{5mm}
We therefore conclude that the  Einstein ``Lanczos Lovelock Gauss-Bonnet"
 equations in the bulk (rather than simply Einstein's)
 induce on the brane a better approximation to
Newton's law and the usual linearized {\it four}-dimensional
 Einstein equations. This is all the more so as the coupling
 parameter $\bar\alpha$ tends to $1$.

Let us be more precise~: at short distances the potential $U(r)$
remains in $1/r$ when $\bar\alpha\neq0$ and hence 
${\cal L}$ needs no longer be smaller than,
 say $200\,\mu\mbox{m}$, 
the upper limit that present experiments~\cite{Ald02} impose
in the $\bar\alpha=0$ case for which $U(r)\propto 1/r^2$. 

As for the coupling constant $G_e\equiv{2+\bar\alpha\over3\bar\alpha}G$
which appears in (7.3) we can identify it to Newton's constant, as
measured in terrestrial laboratories (and known at present with a
$0.15\%$ accuracy, see e.g.~\cite{GunMer00}).

If now, ${\cal L}$ is taken to be of geophysical size (say $\sim 1$ km
$-100$ kms) then the transition from the short to large distance regimes does not
contradict present experiments, whatever the value of $\bar\alpha$
(not too close to zero), as it occurs
``safely'' in a region when gravity is poorly known.

At large distances ($r/{\cal L}\gg1$) the potentials $U(r)\sim V(r)$
are also in $1/r$, with however a coupling constant $G$ which differs from
$G_e$ (identified with Newton's constant as measured in the lab). Now, there is no astro-{\it nomical} evidence that
$G_e$ should be equal to $G$, as only the product $GM$ comes into
play. There is on the other hand some astro-{\it physical} evidence that they
should not be too different, otherwise solar models would go astray.
If we tolerate that $G$ and $G_e$ differ by, say, $10\%$ then
$0.85<\bar\alpha<1.15$.

To summarize : if ${\cal L}$ is of geophysical size, and $0.85<\bar\alpha<1.15$
 (but $\neq1$) then linearized gravity on an Einstein Gauss-Bonnet
 brane is compatible with Newtonian gravity and the light deflection
 experiments (which test the equality $U\sim V$).

We thus see that bulk Einstein Gauss-Bonnet gravity can relax the
constraint on the value of ${\cal L}$ (that is, the characteristic size
of the bulk) at a fairly resasonable price~: $\bar\alpha$ must be
fine-tuned to be not too far from its critical value 
$(1-\bar\alpha)\equiv(1-4\alpha/{\cal L}^2)=\sqrt{1+\alpha\Lambda/3}=0$.

Now, to see if such parameter ranges  can indeed be allowed,
it will be necessary to analyze carefully their implications in the
cosmological context as well~\cite{ChaDuf02}. Investigations in this
direction, however, are beyond the scope of the present paper.

\section*{Acknowledgements}
We aknowledge fruitful correspondence with C. M. Will.
N.D. thanks for its hospitality the Yukawa Institute for Theoretical Physics,
 where this work was performed.
M.S. is supported by the Monbukagaku-sho Grant-in-Aid for Scientific Research (S),
 No.~14102004.

\end{document}